%% file: paper.tex
\RequirePackage[hyphens]{url}
\documentclass[sigconf]{acmart}

\usepackage{lipsum}
\usepackage{amsmath}
\usepackage{graphicx}
\usepackage{comment}
\usepackage{pifont}
\usepackage[frozencache,cachedir=.]{minted}
\usepackage{pgfplots}
\pgfplotsset{every axis legend/.append style={
		at={(0.5,-0.17)},
		anchor=north,legend cell align=left}}
\usepackage[disable]{todonotes}
\usepackage{subcaption}
\newcommand{\sketch}{\todo[color=yellow!30,inline]}  
\usepackage{tikz}
\usetikzlibrary{shapes,snakes}
\usetikzlibrary{patterns}
\usetikzlibrary{plotmarks}
\newcommand*\circled[1]{\tikz[baseline=(char.base)]{
		\node[shape=circle,draw,inner sep=0.5pt] (char) {#1};}}

\newcommand{\pie}[1]{%
	\begin{tikzpicture}
	\draw (0,0) circle (1ex);\fill (1ex,0) arc (0:#1:1ex) -- (0,0) -- cycle;
	\end{tikzpicture}%
}
\begin{document}
\bibliographystyle{acm}
\settopmatter{printacmref=false} 
\renewcommand\footnotetextcopyrightpermission[1]{} 
\pagestyle{plain} 
\settopmatter{printfolios=true}

\title{Evaluating Blockchain Application Requirements and their Satisfaction in Hyperledger Fabric}
\subtitle{A Practical Experience Report}
\author{Sadok Ben Toumia}
\affiliation{\institution{University of Passau}}
\email{bentou01@ads.uni-passau.de}

\author{Christian Berger}
\affiliation{\institution{University of Passau}}
\email{cb@sec.uni-passau.de}

\author{Hans P. Reiser}
\affiliation{\institution{University of Passau}}
\email{hr@sec.uni-passau.de}

\begin{abstract}

Blockchain applications may offer better fault-tolerance, integrity, traceability and transparency compared to centralized solutions. Despite these benefits, few businesses switch to blockchain-based applications. Industries worry that the current blockchain implementations do not meet their requirements, e.g., when it comes to scalability, throughput or latency. Hyperledger Fabric~(HLF) is a permissioned blockchain infrastructure that aims to meet enterprise needs and provides a highly modular and well-conceived architecture.
In this paper, we survey and analyse requirements of blockchain applications in respect to their underlying infrastructure by focusing  mainly on performance and resilience characteristics. 
Subsequently, we discuss to what extent Fabric's current design allows it to meet these requirements. 
We further evaluate the performance of
Hyperledger Fabric 2.2 
 simulating different use case scenarios by comparing single with multi ordering service performance and conducting an evaluation with mixed workloads.
\end{abstract}



\maketitle

\input{sections/intro.tex}

\input{sections/related-work.tex}

\input{sections/background.tex}

\input{sections/method.tex}

\input{sections/requirements-analysis.tex}
\input{sections/hlf-meets-requirements.tex}

\input{sections/evaluation.tex}

\input{sections/conclusion.tex}


\bibliography{literature}

\end{document}

%% file: sections/intro.tex
\section{Introduction}
Since the invention of Bitcoin~\cite{btc}, many people are speculating about how blockchain can revolutionize our daily lives. Several sectors can profit from blockchain, whereas for many other areas it is considered an overkill~\cite{8525392}.
Today, a number of industries still struggle with basic concerns like traceability, integrity protection, or privacy~\cite{pedrothesis,walmartwp,honeywellcasestudy}. 
Competition is higher than ever, which makes certain parties secretive about their transactions. Such issues are often not being sufficiently handled by traditional applications.



This raises the demand of a platform that can handle these issues while meeting their standards in respect to resilience and performance~\cite{honeywellcasestudy,pedrothesis}. 
Enterprises often need a permissioned blockchain that restricts participation to a consortium of members. 
Due to competitors being also on the blockchain network, these parties need privacy: not everyone should be able to see all their transactions -- instead transactions must be on a need-to-know basis~\cite{ibmbenefits}.

Hyperledger Fabric~(HLF)~\cite{androulaki2018hyperledger} is an open-source, permissioned blockchain platform that intends to satisfy enterprise application requirements. It presents a modular architecture with pluggable consensus and can achieve high throughput.
 Previous studies have highlighted the issue of missing support for Byzantine fault tolerance~(BFT)~\cite{DBLP:journals/corr/abs-1709-06921}. Starting from version 2.0, HLF switched from a Kafka-based ordering service
 to a
Raft-based ordering service. While Raft (like Kafka) does not assume BFT, it could be transformed to do so in future, and could essentially be a step towards implementing BFT in HLF. 

We think that it is important to discuss and validate how far design decisions like these, which concern the infrastructure of a blockchain system, match up with the requirements concrete applications impose towards the underlying blockchain infrastructure.


\textbf{Contribution and Outline.}
Our main contribution consists in investigating relevant requirements of blockchain applications and  discussing how far these are addressed in HLF. 
In the remainder of this report, we refer to related work (§\ref{related-work}), provide relevant background knowledge~(§\ref{background}), explain our methodology~(§\ref{methodology}) and conduct a requirements analysis for blockchain applications selected from different use-cases
~(§\ref{requirement-analysis}). Further, we analyse design choices HLF makes to match these requirements~(§\ref{HLF-analysis}) and investigate on the question whether HLF can satisfy performance requirements by conducting experiments for different scenarios~(§\ref{evaluation}). 
Finally, we draw our conclusions~(§\ref{conclusions}).

%% file: sections/related-work.tex
\section{Related Work}
\label{related-work}
\sketch{\~Sadok: Half done?}
Li et al.~\cite{9045815} have recently published a survey paper highlighting Hyperledger Fabric and Hyperledger Composer's use-cases. The paper examined current theoretical and real-life HLF  deployment while highlighting how HLF was used as a solution to solve existing enterprise problems.
A recently published dissertation~\cite{pedrothesis} has studied the requirements and unresolved issues of supply chains while also proposing architectures based on blockchains to address these issues, the main
focus was however restricted to supply chain management.

Several papers have included benchmarks for Hyperledger Fabric~\cite{androulaki2018hyperledger,DBLP:journals/corr/abs-1901-00910,DBLP:journals/corr/abs-1805-11390, baliga2018performance, shalaby2020performance}, mainly focusing
on the \texttt{v1.x} versions of Fabric with FastFabric pushing an impressive 20,000 transactions per second (TPS)~\cite{DBLP:journals/corr/abs-1901-00910}. Androulaki et al. proposed the architecture, components and design choices behind HLF and experimentally validated the system performance~\cite{androulaki2018hyperledger} whereas Thakkar et al. studied how various parameters of the
network impacted performance such as number of channels, number of endorsers and world-state database choice~\cite{DBLP:journals/corr/abs-1805-11390} - their proposals were incorporated in future Hyperledger Fabric versions.

 Further,  Guggenberger et al.~\cite{guggenberger} have recently published a detailed performance report for Hyperledger Fabric v2.0 combining several configurations and testing Fabric's fault-tolerance using DLPS~\cite{sedlmeir2021dlps}. Their report covers an in-depth benchmark analysis of HLF.
 
 In our report, we focus on discussing \textit{application requirements} of blockchain and how these requirements are met by HLF's design. Our report also includes a performance evaluation that conducts experiments on multi ordering service performance and mixed workloads (e.g., mostly read vs. mostly write) which have yet not been sufficiently studied by previous works but are interesting from an application point of view. 
 


\sketch{Todo: to complete.. Review this section.., and add more recent relevant works}

%% file: sections/background.tex
\section{Background}\label{background}


To begin with, we briefly explain important terminology as well as a few fundamental concepts about blockchain and, in particular, Hyperledger Fabric.

\textbf{Blockchain.} 
%
The term blockchain is not used consistently in academic literature. 
Our definition emphasizes that we are referring to a complete system rather than just a specific data structure:

 A \textit{blockchain} is a distributed system that manages an append-only and totally-ordered log of immutable transactions (also called the \textit{ledger}) in a replicated fashion. Several nodes hold a consistent copy of the ledger, and several nodes are involved in validating transactions issued by clients. To order transactions, typically a \textit{consensus} algorithm is employed. Further, transactions are usually grouped into \textit{blocks}, which are chained by referencing the hash of the previous block in the block header.


The traceability and immutable history of transactions in a blockchain fundamentally increase the trustworthiness and transparency of the system.
As long as a sufficiently large portion of nodes in the system (e.g., determined by quantity, resource allocation, or stake) behaves correctly,
the overall system works as intended. There is no need to put trust into the correctness of any single node,
thus eliminating single point of failure for blockchain applications.
Immutability refers to the property that each block bears also the hash of the previous block, and a modification to a block modifies its hash,
which results in the link being broken and thus invalidating subsequent blocks~\cite{DBLP:journals/corr/abs-1709-06921,blockchainzz}.


%

\begin{figure}[tb]
	\centering
	\includegraphics[width=1\linewidth]{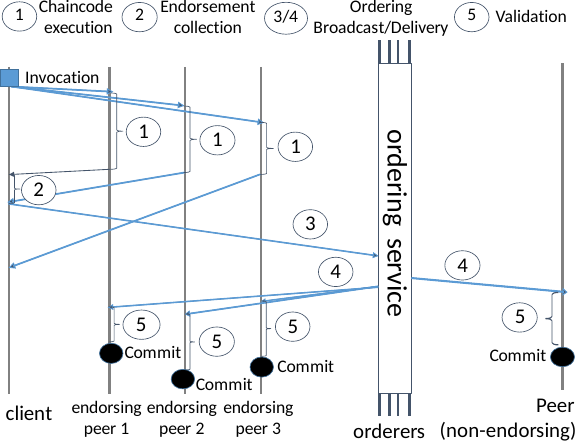}
	\caption{Hyperledger Fabric transaction flow~\cite{androulaki2018hyperledger}.}
	\label{fig:transaction-flow}

\end{figure}

\textbf{Hyperledger Fabric (HLF).} 
HLF is a highly modular enterprise-grade distributed ledger platform. HLF has plug and play capabilities that allow it to be suitable for a wide range of use-cases. Further,
HLF follows an \textit{execute-order-validate} architecture, where transactions are first simulated (this means executed against the current state of the ledger) by endorsing peers, ordered by the ordering service, and then committed by committing peers.

The \textit{transaction flow} consists of the following steps (also shown in Figure~\ref{fig:transaction-flow})~\cite{androulaki2018hyperledger}:
The \textit{client} sends a transaction proposal to the peers specified in the endorsement policy and the \textit{endorsing peers} simulate the transaction (\circled{1}), which produces read sets and write sets, without changing the state of the ledger.
After that, the client collects the responses of the endorsing peers (\circled{2}), which contain the read and write sets, checks if the endorsement policy is satisfied, assembles them in an \textit{envelope} and sends that to the \textit{ordering service} (\circled{3}).
Subsequently, the ordering service orders the transaction without knowing the contents of the envelope. Transactions are batched and once one of the conditions for cutting a block is met, the ordering service sends the block to the committing peers (\circled{4}).
Finally, committing peers validate or invalidate the transactions in the block and the block is eventually appended to the \textit{ledger} (\circled{5}).

%% file: sections/method.tex
\section {Methodology}
\label{methodology}


Our evaluation approach covers two dimensions: \textit{performance} and \textit{resilience}. Performance characteristics are  quantitative (e.g., transaction throughput) and are used to assert that the blockchain can handle the application's workload. Resilience characteristics are often qualitative and describe if the blockchain can deliver a certain service quality (e.g., tolerating faults, providing confidentiality).

\textbf{Aspects of Resilience. }
Resilience is a broad term that encompasses many aspects~\cite{berger2021survey}. 
We employ the following aspects in our subsequent analysis:

\underline{Fault Tolerance Coverage (FTC).}
A measure of effectiveness for fault tolerance is \textit{fault tolerance coverage}~\cite{avizienis04basic}: 
it encompasses the \textit{error- and fault-handling coverage},  a measure to capture how many of the occurring faults are actually covered by the fault tolerance mechanism (development faults might restrict the intended fault-handling coverage) and 
the \textit{fault assumption coverage}, which is a measure for reasoning about how closely  assumptions of a fault model actually cover reality.
In our analysis we employ assumption coverage to indicate which type of faults a blockchain can tolerate. 

\underline{Fault Tolerance Proportion (FTP).}
 Fault tolerance proportion is an assumed upper-bound that indicates the ratio of faulty nodes a blockchain can tolerate, to total nodes participating in the system (this property is also sometimes called resilience bound). 
Fault tolerance coverage and proportion are often coupled.

\underline{Membership (Node Authenticity).} In permissioned blockchains a consortium of nodes is defined  
and a mechanism for managing membership is required. Providing membership information and node authenticity is an important feature for blockchains and  blockchain applications might demand the blockchain system to be capable to changing (e.g., expanding) its consortium at run-time.



\underline{Confidentiality.} There are different \textit{types} of application requirements associated with confidentiality depending
whether the content, sender, or other information of a transaction need to be confidential.




\underline{Integrity.} Integrity is a main motivator towards blockchain adoption. Data integrity in blockchains is achieved by the immutability property of the ledger. Undetected tampering is almost infeasible, as hashes can be used to quickly validate for correctness.
We argue that all blockchain applications share the need of this characteristic and will thus not use it in a comparison.

\textbf{Aspects of Performance.}
We consider typical blockchain performance aspects that application might demand, in particular:

	\underline{Scalability.} Number of nodes that can participate in the blockchain system.
	
	\underline{Throughput.} Number of transactions per second (TPS) that can be processed by the blockchain system.
	 
	\underline{Latency.} Time that elapses between a transaction being issued on the client side and being finalized within a block that is appended to the ledger.

\noindent For each aspect, we categorize performance requirements of block-chain applications into three categories: \textit{low}, \textit{medium} and \textit{high} as shown in Table~\ref{tab:performance-categories}.
This categorization is rough and aligns with performance magnitudes of blockchain systems. 
 Achieving low latency is better and thus means a higher requirement towards the blockchain infrastructure. For the other aspects, higher is better.



\begin{table}[t]
	\centering
	\caption{Categorization of performance requirements.}
	\begin{tabular}{|c|c|c|c|}
		\hline
		& \textbf{Low}        & \textbf{Medium}   & \textbf{High}           \\ \hline
		\textbf{Scalability} & \textless 100 nodes & 100 to 1000 nodes & \textgreater 1000 nodes \\ \hline
		\textbf{Throughput}  & \textless 100 TPS   & 100 to 1000 TPS   & \textgreater 1000 TPS   \\ \hline
		\textbf{Latency}     & \textless 3~s   & 3 s to 10 s       & \textgreater 10 s       \\ \hline
	\end{tabular}%

	 \label{tab:performance-categories}
\end{table}
\sketch{hmm maybe low latency being <3s (one second is a really low bound?) Yes. \#Done}

%% file: sections/requirements-analysis.tex
\section{Requirements Analysis}
\label{requirement-analysis}

HLF is currently being used in a number of fields~\cite{usecaselist}, some of them are already in production, but most of them are still in development or proof-of-concept status. These use-cases are high-risk environments with a lot at stake where some parties could be interested in gaining unauthorized access or tampering with the data for personal gain, thus requiring a highly resilient infrastructure. 

In this section,
we analyse use-cases and derive which of the characteristics (Section~\ref{methodology}) are required by which application. We also present a summary in Table~\ref{tab:comparison}.


\subsection{Electronic Voting (EVote)}
 EVote~\cite{ibmevote,ibmevote2} is an open-source proof-of-concept application for holding an electronic election. The app leverages HLF to meet its needs for immutability and traceability, which
 in return reduce election fraud. Smart contracts are used to tally up votes, therefore reducing costs of manual work~\cite{ibmevote2}.
 A voting network to hold an election is a highly adversarial environment that might encourage malicious behaviour of individuals. Therefore, we consider it valuable for such a system to be
 Byzantine fault-tolerant (and to tolerate up to 33\% of participants becoming faulty). The system should be permissioned. It should further provide high confidentiality: when votes need to be checked
 for their validity (to prevent double voting)  they should be untraceable to the voter to prevent any form of coercion. 
Subnetworks could help to enforce a need-to-know policy for different entities involved in the process.

 From a performance standpoint, a high latency is tolerable in such a network, because voting is per user a one-time action. It should however not
 exceed 30 seconds in order to maintain a pleasant user-experience. To maintain such a latency, the system has to have at least a medium throughput, as elections are usually held in a small time period where at
 peak times, 
 many transactions are issued.
  In such a use-case, the scalability of the system has another goal other than being able
 to handle such a traffic and that is transparency and ensuring that not a single entity has more control over the voting process. An approach for this might be to have every election district host a peer node (or more in
 order to avoid a single point of failure) and as such a medium to high scalability becomes a requirement.

\subsection{Supply Chains (IBM Food Trust and GoDirect Trade)}
GoDirect Trade~\cite{honeywellcasestudy} is a practical use case for blockchain technology, offering an online marketplace for aerospace parts.
The traceability feature of blockchain allows users to access the lifecycle of parts and any associated information required by the government.
IBM Food Trust~\cite{ibmfoodtrust,walmartwp} is a project by Walmart, IBM, Nestle, and Unilever aimed at improving traceability of products and all their ingredients to the farms and also to access different data
about the product in order to satisfy customer needs and guarantee the safety of foods~\cite{walmartwp}.
IBM Foodtrust and GoDirect Trade both utilize the immutability and traceability aspects of blockchain, in that both are interested in the history and provenance of items recorded on the immutable ledger.

 From a resilience perspective
of view, IBM Foodtrust could go well with BFT  (and have a resilience bound of 33\%) whereas GoDirect Trade could benefit from using only CFT  (and having a resilience bound of 50\%). Contrary to IBM Foodtrust, where participants 
could bring up their own peers and deploy their own smart-contracts, GoDirect Trade's nodes are in-house~\cite{honeywellcasestudy}. Further, both applications need to be run on a permissioned blockchain, where
participants are granted access based on their status on the market.
 Moreover, 
both systems require high confidentiality, as trade secrets are at stake here as in both networks competitors are present.

Contrary to EVote, where users are usually one-time users, supply chains, due to the globalisation of the markets, are usually comprised of a lot of actors and each one of them uses the network multiple times in
a small time frame~\cite{pedrothesis}. In terms of throughput and latency, GoDirect Trade requires only a medium throughput and a medium latency wheras IBM FoodTrust requires a high throughput and a low latency. This is due to the to the number of incoming transactions where supply chains in the context of food generate a lot more requests than supply chains in the context of aviation. In GoDirect Trade, network clients are not allowed to host
their own nodes. As stated in~\cite{honeywellcasestudy} the system operates five validating nodes, which indicates that low scalability might suffice. In contrast, IBM FoodTrust subscribes are
allowed to host their own nodes, install their own private smart-contracts on private channels to automate transactions, which indicates that it requires a higher
scalability than GoDirect Trade and therefore needs a least medium scalability.

\subsection{Healthcare (Change Healthcare)}
Change Healthcare~\cite{changehealthcarecs} is a company with the aim to modernize the American health system. 
Leveraging HLF the company is able to link providers and payers in a trustful environment in order to facilitate claims.

As an actor in the healthcare industry, Change Healthcare has to be very wary about how data on their network is handled. Providing access to unauthorized persons has serious legal
consequences~\cite{Peterson2016ABA}, which is why Change Healthcare needs very high confidentiality and private ledgers.
Most importantly, transactions, such as financial or patient data, should be on a subnetwork
with only participating entities granted access (need-to-know basis), for example a hospital at which a person was a patient in and the insurance company for claims processing.
 Similarly to GoDirect Trade, Change Healthcare's nodes are in-house and it can benefit from
providing only CFT and having a fault tolerance proportion of up to $\epsilon<50\%$.

The blockchain network has initially run on six nodes in the company's data-center but now they are looking towards expanding to the cloud. As such, Change Healthcare only needs low scalability due to the nodes belonging to it like in GoDirect Trade's case.
Currently the system can process 550~transactions per second (TPS) but the company is aiming for a higher number in near future~\cite{changehealthcarecs}. As such, the blockchain system needs at least medium throughput and works best with low to medium latency to
maintain a satisfactory user-experience.

\begin{table*}[tb]
	\centering
		\caption[Resilience requirements of blockchain applications towards the underlying blockchain infrastructure.]{Summary of requirements of blockchain applications towards the underlying blockchain
		infrastructure with very high(\ding{77}), high (\pie{360}), medium (\pie{180}) and low (\pie{0}) demands. Note that, lower latency is better, and is thus considered a higher demand towards the infrastructure,  e.g., tolerating a higher latency as in EVote means a lower requirement.}
	\resizebox{1.5\columnwidth}{!}{%
		\begin{tabular}{|c||c|c|c|c||c|c|c|c|}\hline
			\  & \multicolumn{4}{|c||}{\textbf{Resilience}}   & \multicolumn{3}{|c|}{\textbf{Performance}}  \\ \hline
			\textbf{Application} & \textbf{FTC} & \textbf{FTP} & \textbf{Membership} & \textbf{Confidentiality} &\textbf{Scalability} & \textbf{Throughput} & \textbf{Latency} \\ \hline
			EVote& BFT & 33\% & Yes & \ding{77}~(sender, content) & \pie{180}& \pie{180}& \pie{0}  \\\hline 
			IBM FoodTrust& BFT & 33\% & Yes& \pie{360}~(content) & \pie{180}& \pie{360}& \pie{360}   \\ \hline
			GoDirect Trade & CFT & 50\%& Yes & \pie{360}~(content) & \pie{0}& \pie{180}& \pie{180}     \\ \hline
			Change Healthcare& CFT & 50\% & Yes& \ding{77}~(sender, content) & \pie{0}& \pie{180}& \pie{180}   \\ \hline
			Visa B2B Connect& BFT & 33\%  & Yes & \pie{360}~(content) & \pie{180}  &  \ding{77} & \pie{360}     \\ \hline
		\end{tabular}%
	}

	\label{tab:comparison}
\end{table*}

\subsection{Banking (VISA B2B Connect)}
VISA B2B Connect~\cite{visab2b} is a project by VISA to facilitate cross-border and cross-currency payments. It leverages HLF to create a secure and trusted network of financial institutions where international transfers do not have to go through intermediate banks, thus drastically reducing both delays and costs.

 The current standard for cross-border cross-currency payments and the main system VISA B2B Connect is challenging is SWIFT, which handles approximately 33.6 million transactions per day. 
VISA B2B circumvents the shortcoming of traditional banking applications by employing an
one-to-many architecture, in which VISA B2B is directly linked to several financial institutions,  therefore intermediaries can be bypassed.
As a result of this centralization and the SWIFT system as a motivator, such a system requires medium scalability, and very high throughput to be capable of handling peak workloads. Typical other banking methods have varying throughput with PayPal having around 450 TPS~\cite{paypaltps} and credit-card companies such as VISA itself require 50,000 TPS~\cite{DBLP:journals/corr/abs-1901-00910}.

This centralization also means VISA's nodes are in-house. However, unlike GoDiectTrade and other companies hosting their nodes in-house VISA should 
employ
BFT along with a FTP of up to 33\%. The nature of this system makes attacks highly rewarding and insider attacks are a legitimate concern, such as if a participant is compromised or participant
himself being dishonest.

%% file: sections/hlf-meets-requirements.tex
\section{How HLF Meets Enterprise Requirements}
\label{HLF-analysis}

In this section, we focus on the design considerations and features of Fabric that allow it to meet performance and resilience requirements of potential use-cases.
 
\subsection{Resilience Requirements}
In the following we highlight Fabric's features, components and design choices while briefly explaining their role in increasing resilience.

\textbf{Blockchain Features.} 
Maintaining integrity of the data is a critical aspect of resilience and a priority for businesses. 
 HLF, being an implementation of blockchain, comes with both immutability and traceability. Data is immutable once appended to the ledger, this way users can insure its integrity~\cite{androulaki2018hyperledger}.

\textbf{Permissions.} 
Fabric is a permissioned blockchain, permissions are maintained by one or more membership service providers (MSP) which use cryptographic identities. Transactions are checked at every step to verify authenticity of requests. This in turn limits unwanted access and increases trust~\cite{androulaki2018hyperledger}. 

\textbf{Channels.} 
Unlike other blockchain implementations, HLF uses \textit{channels}. A channel is a dedicated subnetwork with its own private ledger and a group of channel members that manage a copy of the ledger, thus ensuring that not every peer on the network has access to the ledger, therefore increasing confidentiality.

\textbf{Endorsement Policy.} 
Channel administrators define the endorsement policy,
which specifies which peers (\textit{endorsers}) have to approve a transaction before this is sent to the ordering service. If a client does not fulfil an endorsement policy 
 he has to retry submitting the transaction again~\cite{androulaki2018hyperledger}. An endorsement policy consisting of multiple peers belonging to different organizations would increase transparency and trust in the system, as no single entity is in full control of endorsing transactions. A single point of failure can be avoided by defining a minimum number or percentage of endorsing peers.

\textbf{Consensus.} 
An appealing feature of HLF is pluggable consensus. Older versions of HLF use Kafka + ZooKeeper (ZK), while the current default consensus protocol is Raft.
Both Raft and Kafka+ZK are crash fault-tolerant and not Byzantine fault-tolerant.
Since consensus is pluggable, however, the developer could opt for a BFT ordering service in future as a new BFT consensus library has been proposed for HLF recently~\cite{barger2021byzantine}.
Raft is embedded into HLF and thus enjoys the direct support of the HLF community whereas Kafka+ZK are supported by Apache. From a performance perspective, a published benchmark~\cite{doeshpscale} with v1.4.1 showed
that Raft can be much more efficient. 


\textbf{Resilience of the Execute-Order-Validate Design.} HLF employs an execute-order-validate architecture to  separate these different concerns. A goal of this design is to help withstanding attacks that may target performance degradation or resource exhaustion. 
In particular, this design can help to circumvent bottleneck situations since it allows for transactions to be processed in parallel and by only a subset of nodes. 

\textbf{General Purpose Language based Chaincode.} 
 Moreover, the Fabric team decided not to limit the programming language choice in smart contracts to domain specific languages and allow developers to use general-purpose languages to minimize programming errors as a result of developers having to learn new languages~\cite{androulaki2018hyperledger}. 

\textbf{Peer Gossip.} 
Peer gossip enables peers' ledgers stay in sync by distributing data to other peers on the channel. This aids resiliency in that peers that have gone offline for sometime are able to have synced ledgers and can endorse transactions again after they are back online.

\textbf{Records of Invalid Transactions.} 
All transactions in HLF, in contrast to other blockchain implementations, are recorded on the ledger whether they are valid or invalid. This allows dishonest or malicious users to be detected and black-listed from the network which results in a more secure platform~\cite{androulaki2018hyperledger}.

\textbf{Identity Mixer.} 
HLF supports the use of \textit{identity mixer} (Idemix) to enhance privacy by providing unlinkability and anonymity -- this however, comes with limitations such as not being able to endorse transactions. An Idemix entity (issuer) certifies a user's attributes in form of a digital certificate, users are then able to generate a zero-knowledge proof of possession of a certificate while revealing only what they choose to reveal to a verifier. 

\textbf{Hardware Security Module.} 
HLF supports the usage of hardware security modules (HSM) allowing cryptographic operations like signature generation to be offloaded to them. This has the advantage of letting the HSM manage private keys of peers or orderers, therefore protecting the keys from unauthorized reading.

\textbf{Transport Layer Security.} 
Communication over a HLF network can be secured using TLS. This can be a one-way or a two-way authentication.

\textbf{Private Data Collections.} 
Channels support data privacy by having only organizations on the channel that are allowed to view these transactions.
In cases where a subset of channel members need to conduct transactions between each other while not wanting other channel members to know the contents of these transactions, they could create a new channel. This is however associated with a higher administrative overhead. A solution for this would be the usage of \textit{private data collections}\footnote{See \url{https://hyperledger-fabric.readthedocs.io/en/release-2.2/private-data/private-data.html}, last accessed 12-22-2020}. Private data has a separate transaction flow compared to other data on the channel.
Only authorized peers can see private data and it is communicated between them using gossip, all other nodes including ordering nodes only see hashes of this data, non-authorized nodes append the hashes of this private data in their ledgers, so they  know a transaction has taken place privately between entities on the channel but they  do not know its content.
To comply with government regulations, some organizations might need to delete private data after a certain time, this is doable and will leave behind a hash in the peer's ledgers as evidence that some data was there~\cite{pvtDatapurge}.

\textbf{Chaincode Lifecycle.} 
Introduced in \texttt{v2.0}, the new Fabric chaincode lifecycle requires organizations participating in the endorsement process to approve a transaction. Previously, in \texttt{v1.x} one organization would define attributes of a chaincode and other organizations choose either to opt-in by installing the chaincode or opt-out and not be able to endorse transactions. The chaincode lifecycle provides equality on a channel by allowing the chaincode to be instantiated only after gathering enough approvals. Chaincode packages also do not need to be identical anymore, different organizations can install different chaincode packages and introduce organization-specific behaviour (for example perform different validations for their interests). This does not conflict with transaction approval as long as endorsement results match~\cite{whatsneww}.

\subsection{Performance Requirements}
Further, some of Fabric's design choices were made to increase the performance of the overall infrastructure.

\textbf{The Advantage of Execute-Order-Validate.} 
In HLF execution and ordering of transactions are separated. This allows for better scalability for both phases while increasing modularity and performance because of the decreased amount of work a node has to do~\cite{androulaki2018hyperledger}.
Some blockchain implementations use an order-execute architecture,  but this design has its limitations.
HLF uses an execute-order-validate approach to allow for parallel execution and eliminate non-determinism of smart contracts (transactions can be processed by a subset of endorsers) therefore increasing throughput and decreasing latency~\cite{androulaki2018hyperledger}.

\textbf{How Channels Help Performance.} 
Dividing the network into channels where each channel is serving a purpose and linking a subset of the organizations on the network while having their own endorsers can increase performance due to the decreased workload. This is HLF's version of sharding (HLF can scale up horizontally using channels), which has frequently been proposed to increase performance in blockchains~\cite{10.1145/3243734.3243853,10.1145/3299869.3319889,doeshpscale}. Generally, the idea of parallelizing transaction processing is an important scalability technique~\cite{berger2018scaling}.

\textbf{Peer Gossip.} 
The optional peer gossip feature of Fabric allows for better performance. The throughput of the ordering service is limited by the network capacity of its nodes, and adding more nodes could decrease throughput. This service elects a leader per organization that pulls blocks from the ordering service and distributes them to the rest~\cite{androulaki2018hyperledger}. This reduces the workload of the ordering service.

\textbf{BatchSize and BatchTimeout.} 
The ordering service in Fabric uses batching and forms blocks out of transactions. A new block is created if
(1)~the number of transactions in the block is equal to the maximum allowed,
(2)~the block's size in bytes has reached max, or
(3)~an amount of time has passed since the first transaction of a new block was received~\cite{androulaki2018hyperledger}.
The parameters \textit{BatchSize} and \textit{BatchTimeout} are customizable, allowing adaptation to the use case. If, however, the wrong values are chosen Fabric's performance can be heavily affected~\cite{hua2020reasonableness}.

\textbf{Supporting Multiple Ordering Services.} 
The ordering service is usually responsible for multiple channels. As the number of channels grows the load on the ordering service grows, scaling the ordering service leads to a performance decrease~\cite{androulaki2018hyperledger}. In cases where adding more channels would overwhelm the ordering service, a new ordering service instance can be brought up%
~\cite{doeshpscale}.

\textbf{World-State Database Choice.} 
Recent work~\cite{cdbldb} has investigated the difference in performance between the supported world-state databases in Fabric. Mostly with lower BatchSizes, LevelDB has shown better performance than CouchDB, but CouchDB offers better functionality through \textit{rich queries}\footnote{See \url{https://hyperledger-
	fabric.readthedocs.io/en/release-2.2/couchdb_as_state_database.html}, last accessed 12-22-2020}. Applications should again make trade-offs here of whether they want more functionality in a database or a better performance. FastFabric has experimented with an in-memory hash table as a ledger~\cite{DBLP:journals/corr/abs-1901-00910} and achieved a huge increase in throughput (from 3200 to 7500 TPS).

%% file: sections/evaluation.tex
\section{Performance Evaluation}
\label{evaluation}

\begin{figure*}[t]
\centering
\input{tikzpictures/multiordering.tex}
\caption{Evaluating the performance impact of ordering services setups in HLF with variabel number of channels.}
\label{fig:multi_ordering}	
\end{figure*}
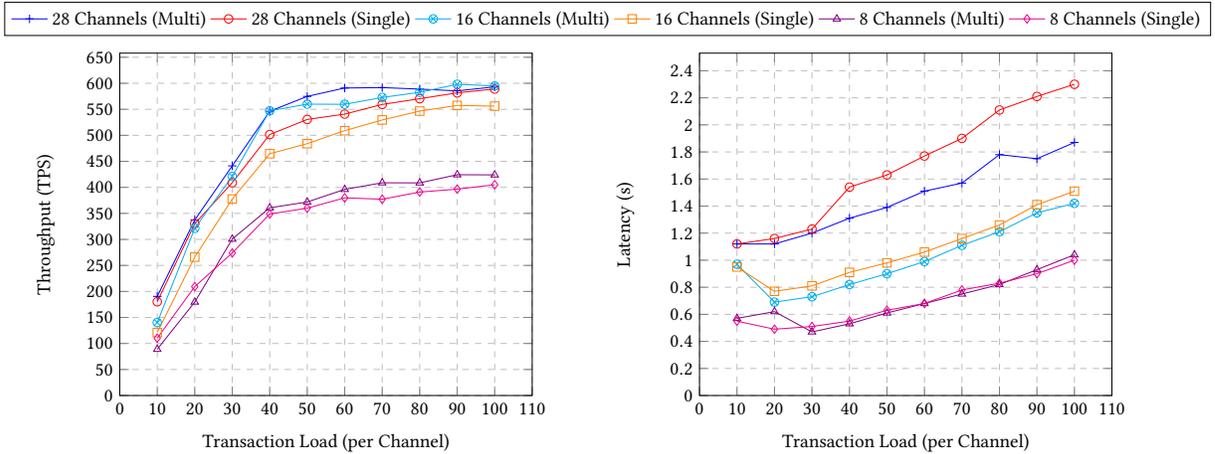

In this section, we aim to examine Hyperledger Fabric's performance under different conditions similar to real-world use-cases in terms of setup and transaction loads. For our purposes, we use Hyperledger's Caliper\footnote{See \url{https://www.hyperledger.org/use/caliper}}, a state of the art tool for benchmarking different blockchain platforms such as Hyperledger Fabric and Ethereum.

In the first part of our evaluation, we focus on the ordering service by examining the benefits of operating a secondary ordering service whilst scaling the number of channels as well as the transaction load.
In the second part of this evaluation, we investigate how HLF performs under different mixed application workloads in terms of read and write operations, thereby simulating real-world scenarios. To mimic a real application, we exemplary choose a \textit{fabcar} chaincode deployment. The \textit{fabcar} is a simple 
chaincode that allows users to add or change data (to be concrete: cars and their ownership) on the ledger using the Fabric contract API\footnote{See \url{https://github.com/hyperledger/fabric-samples/tree/master/chaincode/fabcar}}.
This way we can observe the effect of concurrent reads and writes, i.e., users browsing listings and at the same time users creating new listings. 

For our purposes, we may employ setups with an increasing number of nodes. We are running each node on a 4 vCPU, 6GB RAM Debian VM running in a private OpenNebula cloud and our university's virtualization farm.

\subsection{Multi Ordering Services Performance}

We first examine the benefits of operating a secondary ordering service. The need for a secondary ordering service could arise when the first ordering service is already operating at a high load and servicing a high number of channels or to include only a certain subset of organizations in the ordering phase for certain channels.

\subsubsection{Setup}

We experiment with 8, 16 and 28 two-peer channels where each peer is a member of 2, 4 or 7 channels respectively. 
\begin{minted}[gobble=4,frame=single, fontsize=\small]{yaml}
	   Organizations: 8
	   Peers per Organization: 1
	   Peers per Channel: 2
	   Nodes per Ordering Service: 3
		  MaxMessageCount: 10
		  BatchTimeOut: 0.5
		  Endorsement: 50%
		  Chaincode: fabcar
\end{minted}
For load generation, we used a suitable number of workers for each workload, since employing too many workers can result in inaccuracies in terms of maintained transaction load, while too few workers may not be able to maintain the desired load.
\sketch{\~ Sadok: should i describe caliper as well here? this was done in subsubsection 7.1.2. I dont think the number of workers is relevant here (i used a different number of workers for each txLoad because for ex 2-5 Workers cant really maintain a txLoad of 2800 they are however ok for having a txload of 500. But too many workers, will cause the rate controller to be inaccurate. Since workers are independant of each other, they dont know how many Txs Worker X has pending. They will only generate Txs or back-off based on how many Txs they have pending not on the sum of pending Txs.) \#Done}

\subsubsection{Method} 

In this experiment we scale up the number of channels while experimentally controlling the transaction loads with Hyperledger Caliper.  Note that, for a a multi ordering service setup, each orderer owns half the channels and processes and as such half the transaction load.

Further, the load generation and performance measurement is performed using Hyperledger Caliper where each invocation of the \emph{submitTransaction()} method generates a new transaction per channel thus guaranteeing a fair load distribution among channels and ordering services.


Hyperledger Caliper provides a set of rate controllers to conduct different types of experiments. For our purposes, we decided on the \textit{fixed-load} controller which we slightly modified because it was highly inaccurate when it comes to maintaining a constant load or a minimally-oscillating load. For this, we have overestimated the perceived network throughput in the controller which minimized the delta between the specified \textit{transactionLoad} and the actual load at any time during the experiment.
\subsubsection{Observations}
We make the following observations:

\textit{Observation 1:} Figure 2 (a) shows that throughput is continuously increasing as the transaction load increases and converges to approximately 600 TPS for 28 and 16 channel deployments and to 425 TPS for 8 channel deployments. Increasing the number of channels increases throughput, a 100 Requests per Second (RPS) transaction load per channel achieves approximately 425 TPS for 8 channel deployments and 600 TPS 16 channel and 28 channel multi ordering service deployments. The same holds for a network level load, a 800 RPS network load (50 RPS per channel for a 16 channel deployment and 100 RPS per channel for an 8 channel deployment) achieves approximately 550 TPS at 16 channels compared to 425 TPS at 8 channels.
\sketch{define the term RPS before using it the first time. I guess there i some bug here. "600 TPS and 600 TPS". \#Done

P.S.: no bug, exact values for 16 and 28 channel deployments were 594.4 TPS and 593.1 TPS respectively. i rephrased the sentence to avoid confusion :) }

\textit{Takeaway 1:} Increasing the number of channels increases throughput. The difference in throughput between a 28 channel and 16 channel setup is insignificant with both setups reaching a peak of approximately 600~TPS whereas for an 8 channel setup the peak is reached at approximately 425~TPS. Increasing transaction load also increases throughput however the throughput converges after a certain point.

\textit{Observation 2:} Figure 2 (b) shows that increasing the number of channels leads to an increase in latency. For 28 channel setups the latency difference between a single ordering service setup and a multi ordering service setup is somewhat significant with approximately 2.4~$s$ and 1.8~$s$ respectively.

\textit{Takeaway 2:} Increasing the number of channels increases latency. This increase is more noticeable in single ordering service setups. Latency has increased by more than 100\% between 28 channel and 8 channel setups.

\textit{Observation 3:} Having a secondary ordering service results in a small throughput increase (<20 TPS) and a slight latency improvement.

\textit{Takeaway 3: }A multi ordering service setup does not seem to lead to a significant performance increase.


\begin{figure*}[t]
	\centering
	\input{tikzpictures/mixedworkloads.tex}
	\caption{Evaluating the performance of HLF with different mixed read/write application workloads.}
	\label{fig:eval_workloads}
\end{figure*}

\subsection{Mixed Workloads}

In this experiment we investigate how HLF performs with mixed application workloads. For this reason, we measure performance for different read-to-write ratios, in particular  \textit{mostly write} (20/80 read/write), \textit{equal usage} (50/50 read/write) and \textit{mostly read} (80/20 read/write).

\subsubsection{Setup}

Our mixed workloads deployment, is similar to our multi ordering-service deployment in terms of number of organizations and endorsement. We are using five ordering-service nodes for this deployment since this is a more suitable option in practice.
\begin{minted}[gobble=4,frame=single, fontsize=\small]{yaml}
	   Channels: 1
	   Organizations: 8
	   Peers per Organization: 1
	   Peers per Channel: 8
	   Ordering Service Nodes: 5
	   MaxMessageCount: 100
	   BatchTimeOut: 0.4
	   Endorsement: 50%
	   Chaincode: fabcar
\end{minted}

\subsubsection{Method} 
We evaluate the performance for increasing input rates for which the system is under a transaction load of 100 requests to 1000 requests at any given time depending on the setup. Each invocation of the \emph{submitTransaction()} method results in the generation of a single read or write transaction with a certain probability, e.g., for 20/80 read-write ratio, the probability that a read operation is generated equals  20\%.\newline

\subsubsection{Observations} We make the following observations:

\textit{Observation 1:} The mostly-read workload achieves the highest throughput. The difference between a mostly-write and a mostly-read workload is significant with approximately 120 TPS difference. Equal usage achieves a decent throughput of about 300 TPS, i.e., a 50 TPS increase compared to a mostly-write workload and a 70 TPS decrease compared to a mostly-read workload at 1000 RPS.

\textit{Takeaway 1:}
The mostly-read workload results in a noticeable throughput increase when compared to a mostly-write workload with 376 TPS and 248 TPS respectively at 1000 RPS.

\textit{Observation 2:} Latency increases with an increased transaction load. The mostly write workload achieves the worst latency with approximately 4$s$ at 1000 RPS. Note that the difference between the latencies of the individual workloads is more noticeable at higher transaction loads.
\sketch{Todo, explain the second diagram \#Done}

\textit{Takeaway 2:} An increased RW ratio results in a latency decrease with approximately 4$s$ at a mostly-write workload compared to approximately 2.5$s$ at a mostly-read workload.

\subsection{Discussion}
The obtained results indicate that Hyperledger Fabric achieves performance of several hundreds of transactions per second even on commodity hardware. It is performance-wise superior to some other blockchain platforms, e.g., Ethereum (as of time of writing).

Applying our results to the aforementioned applications, it seems that Fabric meets their requirements to a certain extent. For GoDirect Trade and Change Healthcare, HLF proves to be a perfect fit as a platform. 
For other applications such as Visa B2B Connect and EVote, Fabric lacks BFT support which is vital in these adversarial environments. We assume this aspect will likely change in future releases of HLF.
 Further, for VISA B2B Connect and payment settlement in general, HLF could be a bit slow due to the massive workload (in particular of peak loads) such applications bear.
 
Summarizing, HLF, compared to other solutions, already meets most business requirements performance and security-wise with some trade-offs, and future releases could potentially narrow the gap between enterprise requirements and HLF, especially the planned introduction of BFT.

%% file: tikzpictures/multiordering.tex
	\begin{subfigure}[b]{0.9\columnwidth}
		\centering
		\begin{tikzpicture}[scale=0.8]
		\begin{axis}[
		xlabel={Transaction Load (per Channel)},
		ylabel={Throughput (TPS)},
				grid = both,
		grid style=dashed,
		xmin=0,
		xtick distance=10,
		ymin=0,
		legend style={at={(-0.28,1.1)},anchor=west, legend columns=-1},
		ytick distance=50,]
		\addplot[color=blue, mark=+] table [
		x=tx,
		y=throughput,
		col sep=comma,
		]
		{CSVs/28ChannelsMulti.csv};	
		\addplot[color=red, mark=o]  table [
		x=tx,
		y=throughput,
		col sep=comma,
		]
		{CSVs/28ChannelsSingle.csv};
		\addplot[color=cyan, mark=otimes] table [
		x=tx,
		y=throughput,
		col sep=comma,
		]
		{CSVs/16ChannelsMulti.csv};	
		\addplot[color=orange, mark=square]  table [
		x=tx,
		y=throughput,
		col sep=comma,
		]
		{CSVs/16ChannelsSingle.csv};
		\addplot[color=violet, mark=triangle] table [
		x=tx,
		y=throughput,
		col sep=comma,
		]
		{CSVs/8ChannelsMulti.csv};
		\addplot[color=magenta, mark=diamond]  table [
		x=tx,
		y=throughput,
		col sep=comma,
		]
		{CSVs/8ChannelsSingle.csv};	
		\legend{28 Channels (Multi), 28 Channels (Single),16 Channels (Multi), 16 Channels (Single),8 Channels (Multi), 8 Channels (Single)}

		\end{axis}
		
		\end{tikzpicture}
		\caption{Throughput comparison of multi ordering services and single ordering service setups.}
	\end{subfigure}
	\begin{subfigure}[b]{0.9\columnwidth}
		\centering
		\begin{tikzpicture}[scale=0.8]
		\begin{axis}[
		xlabel={Transaction Load (per Channel)},
		ylabel={Latency (s)},
				grid = both,
		grid style=dashed,
		xmin=0,
		ymin=0,
		xtick distance=10,
		ytick distance=0.2,]
		\addplot [color=blue, mark=+]  table [
		x=tx,
		y=latency,
		col sep=comma,
		]
		{CSVs/28ChannelsMultiLatency.csv};	
		\addplot[color=red, mark=o] table [
		x=tx,
		y=latency,
		col sep=comma,
		]
		{CSVs/28ChannelsSingleLatency.csv};
		\addplot[color=cyan, mark=otimes] table [
		x=tx,
		y=latency,
		col sep=comma,
		]
		{CSVs/16ChannelsMultiLatency.csv};
		\addplot[color=orange, mark=square] table [
		x=tx,
		y=latency,
		col sep=comma,
		]
		{CSVs/16ChannelsSingleLatency.csv};
		\addplot[color=violet, mark=triangle] table [
		x=tx,
		y=latency,
		col sep=comma,
		]
		{CSVs/8ChannelsMultiLatency.csv};
		\addplot[color=magenta, mark=diamond] table [
		x=tx,
		y=latency,
		col sep=comma,
		]
		{CSVs/8ChannelsSingleLatency.csv};
		\end{axis}
		\end{tikzpicture}
		\subcaption{Latency comparison of multi ordering services and single ordering service setups.}
	\end{subfigure}

%% file: tikzpictures/mixedworkloads.tex
	\begin{subfigure}[b]{0.9\columnwidth}
		\centering
		\begin{tikzpicture} [scale=0.8]
		\begin{axis}[
		xlabel={Transaction Load},
		ylabel={Throughput (TPS)},
		xmin=0,
		ymin=0,
		legend style={at={(0.55,0.18)},anchor=west, legend columns=1},
		grid = both,
		grid style=dashed,
		xtick distance=200,
		ytick distance=50,
		]
		\addplot[color=blue, mark=+] table [
		x=tx,
		y=throughput,
		col sep=comma,
		]
		{CSVs/2080Reports.csv};	
		\addplot[color=red, mark=o] table [
		x=tx,
		y=throughput,
		col sep=comma,
		]
		{CSVs/5050Reports.csv};	
		\addplot[color=green, mark=square] table [
		x=tx,
		y=throughput,
		col sep=comma,
		]
		{CSVs/8020Reports.csv};	
		\legend{mostly write, equal usage, mostly read}
		\end{axis}
		\end{tikzpicture}
		\subcaption{Throughput comparison of mixed workloads.}
	\end{subfigure}
	\begin{subfigure}[b]{0.9\columnwidth}
		\centering
		\begin{tikzpicture}[scale=0.8]
		\begin{axis}[
		xlabel={Transaction Load},
		ylabel={Latency (s)},
		xmin=0,
		ymin=0,
		legend style={at={(0.55,0.18)},anchor=west, legend columns=1},
		grid = both,
		grid style=dashed,
		xtick distance=200,
		ytick distance=0.5,]
		\addplot[color=blue, mark=+] table [
		x=tx,
		y=latency,
		col sep=comma,
		]
		{CSVs/2080ReportsLatency.csv};	
		\addplot[color=red, mark=o] table [
		x=tx,
		y=latency,
		col sep=comma,
		]
		{CSVs/5050ReportsLatency.csv};	
		\addplot[color=green, mark=square]  table [
		x=tx,
		y=latency,
		col sep=comma,
		]
		{CSVs/8020ReportsLatency.csv};
		\legend{mostly write, equal usage, mostly read}
		\end{axis}
		\end{tikzpicture}
		\subcaption{Latency comparison of mixed workloads.}
	\end{subfigure}

%% file: sections/conclusion.tex
\section{Conclusion}
\label{conclusions}

Enterprises have previously had minimal interest in blockchains due to the scalability and performance issues. This is however continuously changing in the recent years. The use cases discussed in this paper, as shown in Section~\ref{requirement-analysis}, all have different needs  which make the modularity, customizability, privacy features and the coinless nature of Hyperledger Fabric very attractive.

HLF on the other hand, tries to meet these needs by mainly diverting from traditional architectures like order-execute and by increasing privacy through the usage of channels and private data collections. Its design also allows it to be integrated easily in the way that a potential user could integrate their own certificate authority or employ their own version of an ordering service.

Previous work and our own experience with Fabric have shown that it is massively improving and progressing towards being more decentralized while setting new performance and security standards for other blockchain platforms. The next releases will maybe provide Byzantine fault-tolerance which which would be a major step towards being a production-ready blockchain for high-profile businesses deployed in possibly adverse environments.

\section*{Acknowledgements}
This work has been funded by the Deutsche Forschungsgemeinschaft (DFG, German Research Foundation) grant number 446811880 (BFT2Chain).